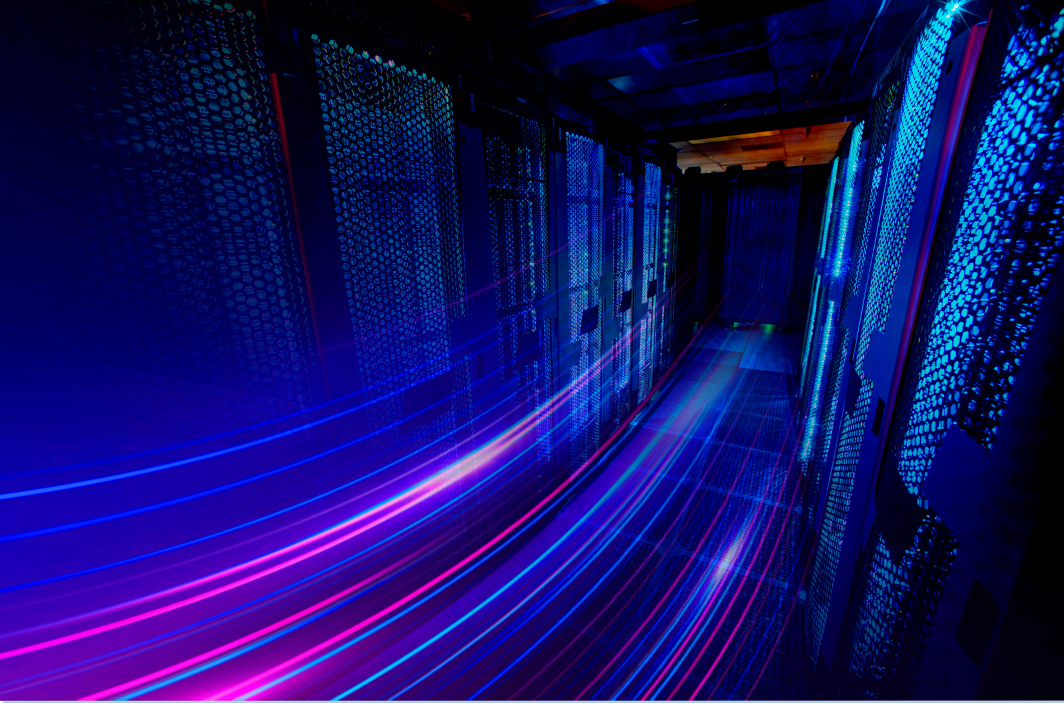

# The Future of International Data Transfers:

## Managing New Legal Risk with a 'User-Held' Data Model

*By* Paulius Jurcys
Marcelo Corrales Compagnucci
Mark Fenwick

# The Future of International Data Transfers:

Managing Legal Risk with a 'User-Held' Data Model

## Paulius Jurcys
Co-Founder, Prifina Inc., Lecturer, Vilnius University Law Faculty
pjurcys@gmail.com

## Marcelo Corrales Compagnucci
Associate Professor of IT Law, Centre for Advanced Studies in Biomedical Innovation Law (CeBIL), Faculty of Law, University of Copenhagen
marcelo.corrales13@googlemail.com

## Mark Fenwick
Faculty of Law, Kyushu University
mdf0911@gmail.com

**Acknowledgment**

The research for this paper was supported by a Novo Nordisk Foundation grant for a scientifically independent Collaborative Research Program in Biomedical Innovation Law (grant agreement number NNF17SA0027784).

# Table of Contents

## Abstract

The General Data Protection Regulation (GDPR) contains a blanket prohibition on the transfer of personal data outside of the European Economic Area (EEA) unless strict requirements are met. The rationale for this provision is to protect personal data and data subject rights by restricting data transfers to countries that may not have the same level of protection as the EEA. However, the ubiquitous and permeable character of new technologies such as cloud computing, and the increased inter-connectivity between societies, has made international data transfers the norm and not the exception. The *Schrems II* case and subsequent regulatory developments have further raised the bar for companies to comply with complex and, often, opaque rules.

    Many firms are, therefore, pursuing technology-based solutions in order to mitigate this new legal risk. These emerging technological alternatives reduce the need for open-ended cross-border transfers and the practical challenges and legal risk that such transfers create post-*Schrems*. This article examines one such alternative, namely a user-held data model. This approach takes advantage of 'personal data clouds' that allows data subjects to store their data locally and in a more decentralised manner, thus decreasing the need for cross-border transfers and offering end-users the possibility of greater control over their data.

## Keywords

GDPR, cross-border data transfers, new Standard Contractual Clauses, personal data clouds, user-held data model, *Schrems II*

# 1 Introduction

The exponential growth of networked technologies, particularly cloud computing, has created an environment in which the large-scale transfer of data, including personal data, is routine. Crucially, such data transfers involve servers located in different countries. Today, many businesses and other organisations engage in cross-border transfers of personal data due to the increasing use of third-party tools, platforms, and products.[1] The use of cloud-based tools for daily information storage, communication, and work, sending emails via automated marketing platforms, and contracting individuals across the globe for customer support and sales are only a few examples of activities that can trigger the application of legal rules on international data transfers.[2] Such transfers are problematic in a European context as the General Data Protection Regulation (GDPR) provides a blanket prohibition to transfer personal data outside the European Economic Area (EEA), unless specific and strict requirements are met.[3] This prohibition, rooted in the previous Data Protection Directive,[4] is certainly contentious and, in today's globalised and interconnected world, spawns legal ambiguities and jurisdictional uncertainties.

These difficulties were exacerbated by the Court of Justice of the European Union (CJEU) in Case C-311/18 Data Protection Commissioner v Facebook Ireland Limited, Maximilian Schrems (*Schrems II*)[5] and its aftermath, which has significantly raised the bar for technology companies looking to meet the resulting regulatory requirements. Transatlantic data transfer underpins more than $1 trillion in cross-border commerce every year and is at the forefront of ongoing trade negotiations between the EU and US.[6]

---

[1] Recitals 6 and 101 of the GDPR; Christopher Kuner, Chapter V 'Transfers of Personal Data to Third Countries or International Organisations' (Articles 44-50),Article 4. 'General Principle for Transfers' in Christopher Kuner, Lee Bygrave and Christopher Docksey (eds) *The EU General Data Protection Regulation (GDPR): A Commentary* (OUP 2020) 757.

[2] Marcelo Corrales Compagnucci, *Big Data, Databases and Ownership Rights of Data in the Cloud* (Springer 2019) 4-7.

[3] Timo Minssen, Claudia Seitz, Mateo Aboy and Marcelo Corrales Compagnucci, The EU-US Privacy Shield Regime for Cross-Border Transfers of Personal Data under the GDPR: What are the legal challenges and how might these affect cloud-based technologies, big data, and AI in the medical sector? European Pharmaceutical Law Review 4(1): 34-38.

[4] Directive 95/46/EC of the European Parliament and of the Council of 24 October 1995 on the protection of individuals with regard to the processing of personal data and on the free movement of such data [1995] OJ L281/31.

[5] C-311/18, *Data Protection Commissioner v. Facebook Ireland Limited, Maximilian Schrems (Schrems II)*.

[6] Most recently, after this paper was submitted for peer review, the EU and US anounced the 'Trans-Atlantic Data Privacy Framework' that should provide durable and reliable legal basis for data flows to foster inclusive and competitive digital economy and lay the foundation for further economic cooperation. See White House,

For this reason, in this paper, we explore an emerging technology-driven alternative that allows individuals to store their user-generated data in their own personal clouds. This approach is a core element of a so-called 'user-held data' approach to user-generated data that gives individuals more control *over* their data and the opportunity to extract more value *from* their data.⁷ One advantage of such an approach is that it greatly reduces the need for unnecessary cross-border data transfers since individuals can choose their geographical location and keep their data locally within the EU, thus mitigating the legal risk for businesses.

After this introduction, Section 2 reviews the main findings of the *Schrems* court decisions and more recent post-*Schrems* regulatory developments. Section 3 suggests that business has adapted to this complex and shifting landscape by doubling down on technological alternatives that decrease the need for cross-border transfers. We focus on the example of user-generated wearable data and how it generates value for individuals and other stakeholders in the 'user-held data model'. The 'user-held data model' allows data subjects to store their data locally and in a more decentralised manner, thus reducing the need for cross-border transfers and offering end-users the tantalising possibility of greater control over their data and opening new opportunities for businesses to easily build applications that run on top of user-held data. Section 4 concludes.

---

'Fact Sheet: United States and European Commission Announce Trans-Atlantic Data Privacy Framework' available at:
<https://www.whitehouse.gov/briefing-room/statements-releases/2022/03/25/fact-sheet-united-states-and-european-commission-announce-trans-atlantic-data-privacy-framework/> accessed 8 April 2022.

⁷ Paulius Jurcys, Christopher Donewald, Mark Fenwick, Markus Lampinen, Vytautas Nekrosius, Andrius Smaliukas, 'Ownership of User-Held Data: Why Property Law is the Right Approach' (2021) Harv JL & Tech Digest, available at:
<https://jolt.law.harvard.edu/digest/ownership-of-user-held-data-why-property-law-is-the-right-approach> accessed 15 January 2022.

## 2 Cross-Border Data Transfers & New Legal Risk

### 2.1 Transnational Data Transfers and the GDPR

One of the main objectives of the GDPR[8] is to guarantee the free flow of personal data between the European Union (EU) Member States. At the same time, however, cross-border data transfers outside of the EEA are forbidden unless the enumerated legal grounds are fulfilled. First, data must be lawfully collected and processed.[9] This includes complying with the principles of purpose limitation[10] and data minimisation,[11] which establish that the purpose of the transfer must be compatible with the one for which the data was collected and limited to what is necessary for that purpose. Second, data exporters and importers must comply with one of the data transfer mechanisms set out in Chapter V of the GDPR.

As an exception to the rule, the GDPR also envisages the prospect of transmitting personal data to a third country – a country outside of the EEA – provided that the party responsible for such a transfer ensures that the data will enjoy the same level of protection as under European standards. Crucially, the GDPR provisions for international data transfers include onward transfers, for example, from a processor to a sub-processor in a third country outside the EEA.[12]

The GDPR provides several mechanisms to ensure this objective of free movement plus EU standards of protection, namely adequacy decisions, appropriate safeguards, and derogations:

- *Adequacy Decisions.* This is an EU Commission determination based on an assessment that third country laws guarantee the same level of protection as under EU law. The effect of such a decision is that personal data can flow without restrictions or any additional safeguards and are regarded as if they were intra-EU.[13]

---

[8] Regulation (EU) 2016/679 of the European Parliament and of the Council of 27 April 2016 on the protection of natural persons with regard to the processing of personal data and on the free movement of such data, and repealing Directive 95/46/EC [2016] OJ L119/1 (General Data Protection Regulation, GDPR).
[9] Art 6 of the GDPR sets out the criteria for lawful processing such as consent, fulfilling a contract, etc. Article 9 of the GDPR provides separate and higher requirements for the processing of special categories of personal data such as explicit consent in the processing of genetic and health data.
[10] Art 5(1)(b) of the GDPR.
[11] Art 5(1)(c) of the GDPR.
[12] Eduardo Unstaran, 'International Data Transfers' in Eduardo Unstaran (ed) *European Data Protection: Law and Practice* (2nd edn, IAPP) 527.

- *Appropriate Safeguards.* In the absence of an adequate level decision, transfer tools containing 'appropriate safeguards' of a contractual nature can be adopted by relevant parties. These include Standard Contractual Clauses (SCCs), Binding Corporate Rules (BCRs), codes of conduct, certification mechanisms, and ad hoc contractual clauses.[14]
- *Derogations.* These have an exceptional nature and are subject to strict conditions such as occasional or non-repetitive processing activities.[15]

This relatively settled framework was disrupted after the Court of Justice of the European Union (CJEU) decision in *Schrems II*. This decision has triggered a flurry of activity in this space, including European Data Protection Board (EDPB) Recommendations on measures that supplement transfer tools to ensure compliance with the EU level of protection of personal data[16] and the new SCCs for the transfer of personal data to third countries[17] issued by the European Commission (EC) on June 4, 2021. These new developments, particularly the new SCCs, have shifted the bar for data protection in international data transfers and introduced a new degree of uncertainty for businesses and privacy professionals. This section introduces these recent developments and some of the critical challenges facing companies in navigating the contours of the post-*Schrems* legal landscape.

---

[13] cf Art 45 of the GDPR. Adequacy Decisions: How the EU determines if a non-EU country has an adequate level of data protection, available at:
<https://ec.europa.eu/info/law/law-topic/data-protection/international-dimension-data-protection/adequacy-decisions_en> accessed 9 October 2021.
[14] cf Art 46 of the GDPR. The transfer tools may require additional 'supplementary measures' to ensure an essentially equivalent level of protection. See C-311/18 (Schrems II), paras 130 and 133.
[15] cf Art 49 of the GDPR.
[16] EDPB Recommendations on measures that supplement transfer tools to ensure compliance with the EU level of protection of personal data. Adopted on 10 November 2020 [hereinafter EDPB Recommendations 01/2020], available at:
<https://edpb.europa.eu/our-work-tools/our-documents/recommendations/recommendations-012020-measures-supplement-transfer_en> accessed 15 January 2022.
[17] European Commission implementing decision of 4 June 2021 on standard contractual clauses for the transfer of personal data to third countries pursuant to Regulation (EU) 2016/679 of the European Parliament and of the Council, C(2021) 3972 final.

**2.2 The *Schrems* Saga**

Legal debates related to cross-border data transfers are associated with two recent cases decided by the CJEU. In 2011, Austrian national Maximilian Schrems spent a semester as a foreign exchange student at Santa Clara University in California. He took some courses related to law, technology, and data privacy. During those months, Schrems learned about the significance of user data to Silicon Valley technology companies and various inadequacies in the framework of data and privacy rights protection in US and EU law.[18]

In 2013, Schrems filed a legal suit with the Irish Data Protection Commission (DPC) against Facebook (*Schrems I*).[19] He requested a suspension of the transfer of his personal data from Facebook Ireland to the US and a declaration that any such transfer was illegal under EU data protection law. He considered that the law and practice of the US did not provide adequate protection of the personal data held in its territory. In particular, he was concerned about the surveillance activities engaged in by various public authorities such as the US National Security Agency (NSA).[20]

The DPC, however, rejected the complaint on the grounds that Decision 2000/520 (Safe Harbour Agreement) ensured an adequate level of protection of personal data transferred between the EU and the US. Schrems contested the decision of the DPC, and the Irish High Court referred the case to the CJEU for a preliminary ruling on the interpretation and validity of the Safe Harbour Framework.[21]

In addition to a lack of informed consent regarding the use of his data and a failure to provide adequate legal remedies for individuals, Schrems argued that US surveillance laws[22] and US programs disclosed by the Snowden revelations[23] such as PRISM,[24] allowed the US authorities unrestricted access to personal data gathered by US technology companies, including Facebook. The CJEU upheld Schrems' case and decided to overturn the Safe Harbour Agreement in 2015, thereby making the transfer of personal data illegal under this framework. The CJEU held that this was a violation of European privacy laws and the fundamental human rights principles enshrined in Articles 7, 8 and 47 of the Charter of Fundamentals Rights of the European Union (CFR).[25] After the invalidation of the Safe Harbour Agreement, Facebook and other companies responded by adopting SCCs to provide a legal basis for the transmission of personal data outside of the EU to the US.[26]

---

[18] Kashmir Hill, 'Max Schrems: The Austrian Thorn in Facebook's Side' *Forbes* (7 February 2012) available at: <https://www.forbes.com/sites/kashmirhill/2012/02/07/the-austrian-thorn-in-facebooks-side/?sh=5557aa247b0b> accessed 15 January 2022.
[19] C-362/14, *Maximillian Schrems vs Data Protection Commissioner* (*Schrems I*).
[20] Noyb, 'EU-US Data Transfers' available at: <https://noyb.eu/en/project/eu-us-transfers> accessed 15 January 2022.
[21] cf Minssen (n3) 39-41.
[22] Such as S 702 of the Foreign Intelligence Surveillance Act (FISA) and Executive Order 12333.
[23] Barton Gellman, *Dark Mirror: Edward Snowden and the American Surveillance State* (Penguin 2020).
[24] Nicholas Watt, 'Prism Scandal: European Commission to Seek Privacy Guarantees from the US' *The Guardian* (10 June 2013), available at: <https://www.theguardian.com/world/2013/jun/10/prism-european-commissions-privacy-guarantees> accessed 15 January 2022.

In July 2016, and to alleviate concerns related to trans-Atlantic data transfers, the EU Commission adopted a new framework – the 'EU-US Privacy Shield.'[27] The Privacy Shield was based on four main pillars:

- A requirement that companies engaged in such transfers are transparent.
- More significant limitations on companies and the development of more adequate supervision mechanisms.
- More opportunities for legal redress for individuals and more effective legal mechanisms for alternative dispute resolution (ADR).
- An annual review involving both EU and US authorities and other stakeholders. The EU-US Privacy Shield framework included a list of requirements such as submitting a privacy policy with specific details. As such, the new framework required greater transparency, oversight, and redress mechanisms, including the creation of an ombudsman to investigate complaints, as well as arbitration and ADR mechanisms.[28]

The EU-US Privacy Shield was greeted with high hopes and expectations at its inception. For instance, in her press release, Věra Jourová, the former EU Commissioner for Justice,

---

[25] Marcelo Corrales Compagnucci, Timo Minssen, Claudia Setiz and Mateo Aboy, 'Lost on the High Seas Without a Safe Harbor or a Shield? Navigating Cross-Border Transfers in the Pharmaceutical Sector After *Schrems II* Invalidation of the EU-US Privacy Shield' (2020) 4(3) EPLR 153, 154-155.
[26] Leslie Hamilton, 'The Legal Environment' in Leslie Hamilton and Philip Webster (eds) *The International Business Environment* (4th edn, OUP 2018) 341.
[27] Commission Implementing Decision (EU) 2016/1250 of 12 July 2016 pursuant to Directive 95/46/EC of the European Parliament and of the Council on the adequacy of the protection provided by the EU-US Privacy Shield (notified under document C(2016) 4176).
[28] cf Minssen (n3) 38.

Consumer Rights, and Gender Equality, stated that 'the EU-US Privacy Shield is a robust new system to protect the personal data of Europeans and ensure legal certainty for businesses… The new framework will restore consumers' trust when their data is transferred across the Atlantic.'[29]

Private and public companies in various sectors responded by adopting this framework, which allowed firms to legally engage in cross-border data transfers between the EU and the US once again.[30] Nevertheless, the EU-US Privacy Shield was not met with universal enthusiasm. Schrems, himself, commented on Twitter that '[t]hey put ten layers of lipstick on a pig, but I doubt that the Court and DPAs suddenly want to cuddle with it.'[31] Soon after, Schrems instituted new court proceedings – *Schrems II* – challenging the legality of the new EU-US Privacy Shield. The CJEU judgment in this second case was handed down on July 16, 2020. The Court found the EU-US Privacy Shield to be invalid but upheld the validity of Standard Contractual Clauses.[32]

Regarding the EU-US Privacy Shield, the CJEU doubted whether US law effectively ensured an adequate level of protection as prescribed under Article 45 GDPR, relating to the fundamental rights guaranteed by the CFR. In particular, the CJEU considered that US law did not grant the necessary limitations and safeguards regarding interferences authorised by national legislation and did not ensure adequate judicial protection against any such interference.

However, regarding SCCs, the CJEU required a case-by-case analysis on their application. Data controllers and data processors exporting data to third countries should verify whether local laws and practices in the third country impinge on the effectiveness of the appropriate safeguards as contained in GDPR Art. 46. Data controllers are now expected to implement 'supplementary measures' to fill any deficiencies and ensure the level of protection required by EU law. Concerning non-EU organisations importing data from the EU based on SCCs, the CJEU also noted that they must inform data exporters in the EU if they cannot comply with the SCCs. In such cases, the data exporter based in the EU must suspend the data transfer or terminate the contract. However – to the dismay of many commentators – the Court did not specify precisely what these supplementary measures might be.[33]

---

[29] European Commission, 'European Commission launches EU-US Privacy Shield: Stronger protection for transatlantic data flows' (12 July 2016), available at:
<https://ec.europa.eu/commission/presscorner/detail/en/IP_16_2461> accessed 15 January 2022.

[30] For a list of over 5,300 companies relying on this framework see: <https://www.privacyshield.gov/list> last accessed on 15 January 2022.

[31] David Gilbert, 'Safe Harbor 2.0: Max Schrems Calls 'Privacy Shield' National Security Loopholes "Lipstick On A Pig"' (29 February 2016), available at:
<https://www.ibtimes.com/safe-harbor-20-max-schrems-calls-privacy-shield-national-security-loopholes-lipstick-2327277> last accessed on 15 January 2022.

[32] C-311/18 (Schrems II) para 203.

**2.3 After *Schrems II*?**

In response to the *Schrems II* decision, in November 2020, the European Data Protection Board (EPBD) introduced recommendations on measures that 'supplement transfer tools' and ensure compliance with an EU level of personal data protection. The Recommendations were intended to assist companies in putting into practice the requirements established by the CJEU in *Schrems II*.

In particular, the EDPB recommended that organisations adopt six steps in cases involving the transfer of personal data to countries outside of the EEA:[34]

- *Step 1 – Know the Data Transfer.* Data exporters should be fully aware of their transfers of personal data to third countries, including onward transfers. Mapping and recording data transfers can be a challenging task. However, this first step is necessary to ensure an equivalent level of protection wherever the data is subsequently processed. Therefore, data exporters should record all processing activities, ensure data subjects are adequately informed, and make sure transfers follow the principle of data minimisation.[35]
- *Step 2 – Identify the Transfer Mechanism.* This involves identifying the transfer tools set out in Chapter V of the GDPR, i.e., adequacy decision;[36] appropriate safeguards of a contractual nature;[37] and derogations.[38] If a data transfer does not fall under either adequacy decisions or derogations, it is necessary to continue to Step 3.[39]

---

[33] Laura Bradford, Mateo Aboy and Kathleen Liddell, 'Standard Contractual Clauses for Cross-Border Transfers of Health Data After *Schrems II*' (2021) 8(1) Journal of Law and the Bioscience 1-2.
[34] Marcelo Corrales Compagnucci, Mateo Aboy and Timo Minssen, 'Cross-Border Transfers of Personal Data after Schrems II: Supplementary Measures and New Standard Contractual Clauses (SCCs)' (2022) 4(2) NJEL 37-47.
[35] EDPB Recommendations 01/2020 on measures that supplement transfer tools to ensure compliance with the EU level of protection of personal data (hereinafter 'EDPB Recommendations 01/2020') (adopted on 10 November 2020) 8-9.
[36] See Art 45 of the GDPR. Adequacy decisions may cover a country as a whole or be limited to a part of it. If you transfer data to any of these countries, there is no need to take any further steps described in this section. The EU Commission has so far recognised only twelve countries which can offer an adequate level of protection. These countries are Andorra, Argentina, Canada (commercial organisations), Faroe Islands, Guernsey, Israel, Isle of Man, Japan, Jersey, New Zealand, Switzerland and Uruguay. As of March 2021,

- *Step 3 – Assess Whether Art. 46 of the GDPR Transfer Tool is Effective Considering all the Transfer Circumstances.* Utilising a transfer mechanism under Art. 46 of the GDPR may not be sufficient if the transfer tool is not 'effective' in practice. In this context, effective means that the level of protection is essentially equivalent to that afforded within the EEA.[40] Data exporters should carry out a Transfer Impact Assessment (TIA) to assess – in collaboration with the importer – if the law and practice of the third country where the data is being transferred may impinge on the effectiveness of the appropriate safeguards of the Art. 46 in the context of the specific transfer. In performing this assessment, different aspects of the third-country legal system should be considered, particularly whether public authorities can access personal data and, in general, those elements enlisted in Art. 45(2) of the GDPR,[41] such as the rule of law situation and respect for human rights in that third country.[42]

- *Step 4 – Adopt Supplementary Measures.* If the TIA reveals an Art. 46 tool is not effective, data exporters – in collaboration with the importers, where necessary – need to consider if any 'supplementary measures' exist. Supplementary measures are supplementary to the transfer tools' safeguards. In other words, if they are added to the safeguards contained in Art. 46, they ensure that the data transferred is afforded an adequate level of protection in the third country, equivalent to the European standard. Exporters need to identify, on a case-by-case basis, which supplementary measures would be effective considering the previous analysis in steps 1, 2 and 3.[43]

- *Step 5 – Formal Procedural Steps.* Taking formal procedural steps in a case where effective supplementary measures have been identified. The procedure may vary

---

adequacy talks were concluded with South Korea. See European Commission, Adequacy decisions, How the EU determines if a non-EU country has an adequate level of data protection, available at: <https://ec.europa.eu/info/law/law-topic/data-protection/international-dimension-data-protection/adequacy-decisions_en> accessed 15 January 2022.

[37] Art 46 of the GDPR. The transfer tools may require additional 'supplementary measures' to ensure an essentially equivalent level of protection. See C-311/18 (Schrems II), paras 130 and 133.

[38] Art 49 of the GDPR.

[39] EDPB Recommendations 01/2020, 9-11.

[40] C-311/18 (Schrems II) para 105 and the second finding.

[41] C-311/18 (Schrems II) para 104.

[42] EDPB Recommendations 01/2020, 12.

[43] EDPB Recommendations 01/2020, 15-17.

depending on the transfer tool that has been used or is expected to be used. For instance, if data exporters intend to put in place supplementary measures in addition to the SCCs, there is no need to request approval from the supervisory authority (SA), insofar that the supplementary measures do not contradict, directly or indirectly, the SCCs and are adequate to ensure that the level of protection guaranteed by the GDPR is not compromised in any way.[44]

- *Step 6 – Re-Evaluate at Appropriate Intervals.* The final step is to monitor and review, on an ongoing basis, whether there are new developments in the third country where data is transferred, which might affect the initial assessment of the level of protection of the third country and the supplementary measures taken based on the TIA and the specific transfer. This is in line with the principle of accountability, a continuous obligation set out in Art. 5(2) of the GDPR. In collaboration with the importers, data exporters should put in place sound mechanisms to ensure that any transfer relying on the SCCs is suspended or prohibited if the supplementary measures are no longer effective in that third country or where those clauses are breached or become impossible to honour.[45]

Following the EDPB Recommendations, on June 4, 2021, the European Commission issued its long-awaited post-*Schrems II* SCCs to transfer personal data to third countries.[46] SCCs are model provisions that contracting parties can adopt, often as addenda to data processing agreements. They impose security and privacy obligations on all parties in cases involving data transfer to countries that do not have an adequate level of protection.[47] The updated versions of the SCCs reflect the GDPR requirements and the legal position after the *Schrems II* decision. The Commission adopted two sets of SCCs, one for use between

---

[44] EDPB Recommendations 01/2020, 17-18.
[45] EDPB Recommendations 01/2020, 18-19.
[46] European Commission implementing decision of 4 June 2021 on standard contractual clauses for the transfer of personal data to third countries pursuant to Regulation (EU) 2016/679 of the European Parliament and of the Council, C(2021) 3972 final.
[47] Ian Lloyd, *Information Technology Law* (7th edn, OUP 2014) 183.

controllers and processors in the EU/EEA[48] and one for transferring personal data to third countries.[49]

In general terms, the new SCCs have the following features:[50]

- *A Modular Approach.* The new SCCs provide more flexibility for complex processing chains by adopting a so-called 'modular approach.' This means that data exporters and data importers can now choose the module that best applies to their needs within the agreement.[51]
- *Geographic Scope of Application.* The new SCCs have a broader application than the earlier clauses, allowing the data exporter to be a party if it was established inside the EEA. According to the new SCCs, the data exporter can be a non-EEA entity.[52]
- *Multipartite Clauses and Docking Clause.* The new SCCs facilitated multiple data exporting parties to conclude a contract (for instance, within corporate groups or collaborations) and for new parties to be added through a so-called docking clause.[53] This clause is optional and allows other third parties that are not a party to an agreement to subsequently join and be included in an agreement without having to conclude a new and separate agreement.[54]

---

[48] European Commission, Standard Contractual Clauses for controllers and processors in the EU (4 June 2021), available at:
<https://ec.europa.eu/info/law/law-topic/data-protection/publications/standard-contractual-clauses-controllers-and-processors> accessed 15 January 2022.
[49] European Commission, 'Standard Contractual Clauses for international transfers' (4 June 2021), available at:
<https://ec.europa.eu/info/law/law-topic/data-protection/international-dimension-data-protection/standard-contractual-clauses-scc/standard-contractual-clauses-international-transfers_en> accessed 15 January 2022.
[50] cf Corrales Compagnucci et al (n 34) 37-47.
[51] Module 1: controllers to controllers, Module 2: controllers to processors, Module 3: processors to processors and Module 4: processors to controllers. This is, for example, when there is a processor such as a cloud service provider located in the EEA and transfers data to another processor such as an infrastructure provider in the US. In this case, the data is transferred back to the controller (back to 'its origin'). This is also sometimes called 'reverse transfer'.
[52] Phillip Lee, 'The Updated Standard Contractual Clauses: A New Hope?' *IAPP* (7 June 2021) available at: <https://iapp.org/news/a/the-updated-standard-contractual-clauses-a-new-hope/> accessed 15 January 2022.
[53] ibid.
[54] Alexander Milner-Smith, 'New Standard Contractual Clauses – what do you need to know?' (14 June 2021), available at:
<https://www.lewissilkin.com/en/insights/new-standard-contractual-clauses-what-do-you-need-to-know> accessed 15 January 2022; Martin Braun, Kirk Nahra, Frédéric Louis, Shannon Togawa Mercer and Ali Jessani, 'European Commission adopts and publishes new Standard Contractual Clauses for international transfers of personal data' *WilmerHale* (7 June 2021) available at:
<https://www.wilmerhale.com/en/insights/blogs/wilmerhale-privacy-and-cybersecurity-law/20210607-european

- *Data Transfer Impact Assessment (TIA)*. Companies must conduct and document a mandatory TIA and make any such exercise available to the competent supervisory authority upon request. The TIA should assess and warrant the following: whether the third country's laws into which the data is imported is compliant with the SCCs and the GDPR and whether any additional safeguards are necessary to enhance data protection.[55]
- *Security Measures*. Annex II of the new SSCs provides a more detailed list of examples of the technical and organisational measures necessary to ensure an appropriate level of protection, including measures to ensure data security.[56]

It seems clear that the new SCCs have raised the threshold for data protection and security in international data transfers. Adopting and complying with this new legal framework has created additional and substantial legal, organisational, and technical requirements for businesses.

Even a cursory review of these developments reveals the high degree of legal risk that businesses are exposed to by these developments. Moreover, this new legal risk has an uncertain and constantly evolving character that adds to the challenge of managing such risk. Given the centrality of data for many businesses today, however, firms cannot ignore this legal risk but must develop strategies to manage it. In the following section, we explore how emerging technologies might provide one strategy that firms might adopt to reduce the legal risk created by the above regulatory developments.

## 3 Technology-Driven Alternatives

The overall business response to the developments described above has, unsurprisingly, been sceptical, and there are significant concerns regarding the new framework.[57] The *Schrems II*

---

-commission-adopts-and-publishes-new-standard-contractual-clauses-for-international-transfers-of-personal-data> accessed 15 January 2022.

[55] Caitlin Fennessy, 'Data transfers: questions and answers abound, yet solutions elude' *IAPP* (12 February 2021), available at: <https://iapp.org/news/a/data-transfers-questions-and-answers-abound-yet-solutions-elude/> accessed 15 January 2022.

[56] cf Annex II Standard Contractual Clauses.

[57] Angelique Carson, 'Schrems II problem may be a long fix' (12 March 2021) available at: <https://www.osano.com/articles/schrems-ll-problem-may-be-long-fix> accessed 15 January 2022.

decision imposes a considerable burden on many organisations without necessarily increasing the degree of protection given to data subjects. Parties must follow a risk-based approach and be prepared to perform a TIA considering the EDPB Recommendations and new SCCs requirements. Moreover, the high adequacy threshold seems likely to result in delayed or terminated data transfers that are potentially damaging for firms. Large and small companies now have to make substantial organisational, technical, and legal reforms that will require considerable costs and resources. This might not even be feasible for many companies dealing with a complex web of data transfers involving hundreds if not thousands of overseas recipients, such as in the case of a cloud federation or brokerage scenarios. Finally, performing a TIA seems likely to be difficult, costly, and time-consuming for many firms that may lack the capacities or resources to perform such an exercise.

Given the complexities surrounding these developments, there has been a strong push from the private sector for alternative solutions.[58] Here, we suggest that a necessary adaptation to this new data environment has been to double down on solutions that leverage digital technologies to significantly reduce the need for cross-border data transfers. To illustrate this argument, we describe what we characterise as a 'user-held data model' (section 3.1) and show that the adoption of a user-centric, user-held data model where user-generated data is held and processed in individuals' personal data clouds can eliminate the need for cross-border transfers of data and associated legal risks.

## 3.1 A User-Held Data Model

Three factors are providing the impetus for the emergence of technology-driven alternatives, such as a user-held data model. First, the GDPR and similar regulatory developments in other jurisdictions, for example, the California Consumer Privacy Act (CCPA) created new requirements for companies to be more transparent and responsible in managing their customer data.[59] By introducing data portability rights, the regulators aim to give consumers more actual control and agency over their personal data and make user-generated data more accessible to third-parties.[60] The most recent proposals in the EU - most notably, the EU Strategy for Data and EU Data Act - aim to unlock the data from siloes and facilitate the creation of new types of services on top of user-generated data.[61]

---

[58] One of the most notable initiatives in Europe is GAIA-X which is a software framework primarily designed for private organisations and that can be deployed to any existing cloud/edge technology stack to obtain transparency, controllability, portability and interoperability across data and services. See <https://www.gaia-x.eu> accessed 15 January 2022.

[59] In California, such a retrieval of personal data is possible pursuant to Art 1798.110 of the CCPA (right to know) and S 999.312 of the CCPA Regulations.

The second factor contributing to the emergence of technological alternatives relates to the practical problems businesses face regarding customer data. As already alluded to in section 2 before, GDPR and other data privacy regulations imposed high compliance costs upon companies which had to implement various safeguards and procedures to ensure that customer data was secure and used lawfully.[62] Nevertheless, data privacy regulations have not changed the huge asymmetry among businesses: the data titans (Amazon, Apple, Facebook, Google, and Microsoft) dominate the data market.[63] Such asymmetries in the data market and compliance-related considerations have pushed businesses to explore new ways to interact with their customers directly and search for ways to obtain relevant 'first-party' data directly from consumers.[64]

Finally, data processing technologies have reached a level of maturity where decentralised data collection and management models have now become more feasible. In particular, machine learning, differential privacy, edge computing, decentralised ledger (blockchain) technologies and AI make it possible to conduct significant data processing locally (i.e., at the edge of the network or on end-user devices).[65] On the other hand, centralised data models struggle to provide adequate performance and safeguards to advanced applications.[66] This shift towards a decentralised, user-centric, user-held data ecosystem is likely to curtail the risk of significant data leaks because hackers will find fewer incentives to hack data sets stored on independent personal data clouds of each user.[67]

---

[60] Paul De Hert, Vagelis Papakonstantinou, Gianclaudio Malgieri, Laurent Beslay and Ignacio Sanchez, 'The right to data portability in the GDPR: Towards user-centric interoperability of digital services' (2018) 34(2) Computer Law & Security Review 193; Sophie Kuebler-Wachendorff, Robert Luzsa, Johann Kranz, Stefan Mager, Emmanuel Syrmoudis, Susanne Mayr, and Jens Grossklags, 'The Right to Data Portability: conception, status quo, and future directions' (2021) 44 Informatik Spektrum 264.
[61] cf Data Act: Proposal for a Regulation on harmonised rules on fair access to and use of data, <https://digital-strategy.ec.europa.eu/en/library/data-act-proposal-regulation-harmonised-rules-fair-access-and-use-data> accessed 7 April 2022.
[62] cf Office of the Attorney General, 'Standardized Regulatory Impact Assessment: California Consumer Privacy Act of 2018 Regulations' <http://www.dof.ca.gov/Forecasting/Economics/Major_Regulations/Major_Regulations_Table/documents/CCPA_Regulations-SRIA-DOF.pdf> accessed 15 January 2022. According to statistics available in September 2020, the average cost for the processing of a single customer's data request was approximately 1500 US dollars, see Wirewheel, 'The Ultimate Guide to Data Subject Access Request (DSAR) Management' <https://wirewheel.io/wp-content/uploads/2019/10/WireWheel-eBook-Ultimate-Guide-to-Data-Subject-Access-Request-Management.pdf> accessed 29 September 2020, 3.
[63] Nicholas Economides and Ioannis Lianos, 'Restrictions on Privacy and Exploitation in the Digital Economy' (2021) 17(4) Journal of Competition Law & Economics 765.
[64] Jasmin Malik Chua, 'Direct-to-Consumer's Lasting Impact on Fashion' *Vogue Business* (2 February 2020) available at: <www.voguebusiness.com/consumers/direct-to-consumer-lasting-impact-on-fashion-levis-nike-samsonite> accessed 15 January 2022; Kati Chitrakorn, 'Zero-party data: The new marketing frontline for luxury' *Vogue Business* (1 October 2021) available at: <https://www.voguebusiness.com/companies/zero-party-data-the-new-marketing-frontline-for-luxury-brands> accessed 15 January 2022.

A 'user-held data model' is based on the premise that individuals are at the heart of the data ecosystem.[68] Individuals generate the data by their activities either in the real world or online. An obvious example of how individuals create data are wearable devices with sensors measuring location, daily steps, heart rate, and capturing numerous other physical parameters. We use wearables as an illustration in the following section.

A user-held data model can be seen as the evolution and synthesis of previous developments in user-centred design,[69] user-centred UX/UI,[70] privacy-by-design,[71] software development, cloud and edge computing,[72] and data science over the recent two decades. A recurring theme of all these developments is the empowerment of individuals with user-friendly, privacy-preserving tools that grant individuals more personal agency and control over data-driven consumer products and, by extension, their lives, and at the same time help companies shift away from product-centred business models and offer a stickier yet friction-free customer experience.[73]

---

[65] Dijiang Huang, Tianyi Xing, and Huijun Wu, 'Mobile Cloud Computing Service Models: A User-Centric Approach' (2013) IEEE Network (September/October) 6-11; Juan C Yelmo, Jose M Del Alamo, and Ruben Trapero, 'Privacy and Data Protection in a User-Centric Business Model for Telecom Services' in IFIP International Federation for Information Processing, Vol. 262, The Future of Identity in the Information Society (Boston: Springer) 447-461.

[66] Mang Su, Fenghua Li, Guozhen Shi, Kui Geng, and Jibo Xiong, 'A User-Centric Data Secure Creation Scheme in Cloud Computing' (2016) 25(4) Chinese Journal of Electronics 753; Jouko Ahvenainen, 'Digital giants own you now, but could it be very different tomorrow?' *Disruptive Asia* (28 October 2021) available at: <https://disruptive.asia/digital-giants-own-you-now-very-different-tomorrow/> accessed 15 January 2022; Jouko Ahvenainen, 'Who protects whom in the battle for personal data and privacy? You?' *Disruptive Asia* (25 March 2021) available at:
<https://disruptive.asia/who-protects-whom-in-the-battle-for-personal-data-and-privacy-you/> accessed 15 January 2022.

[67] Han Qiu, Hassan Noura, Meikang Qiu, Zhong Ming, and Gerard Memmi, 'A User-Centric Data Protection Method for Cloud Storage Based on Invertible DWT' (2021) 9(4) IEEE Transactions on Cloud Computing 1293; Paulius Jurcys, 'Getting Value with User-Held Data' *Medium* (17 November 2020) available at: <https://medium.com/prifina/getting-value-with-user-held-data-cc7f518d96b4> accessed 15 January 2022.

[68] Jurcys (n7); Corrado Moiso and Roberto Minerva, 'Towards a user-centric personal data ecosystem: The role of the bank of individuals' data' (2012) 16th International Conference on Intelligence in Next Generation Networks 202.

[69] Donald A Norman, *User-Centered System Design: New Perspectives on Human-Computer Interaction* (CRC Press 1986).

[70] Donald A Norman, *Design of Everyday Things* (Basic Books 2013).

[71] Ann Cavoukian, 'Privacy by Design; The 7 Foundational Principles' available at:
<https://www.ipc.on.ca/wp-content/uploads/resources/7foundationalprinciples.pdf> accessed 15 January 2022.

[72] cf '6G White Paper on Edge Intelligence' (June 2020) available at:
<http://jultika.oulu.fi/files/isbn9789526226774.pdf> accessed 7 April 2022.

The user-held data model refers to the technological architecture where individuals can gather and connect various data sources to one's own 'personal data cloud'.[74] Such data could be coming from an individual's online accounts (e.g., one's activity history from YouTube or Netflix) or data collected by wearable activity trackers and IoT devices (such as smartwatches, smart rings, smart home appliances, etc.). A personal data cloud is a central repository where individuals can pool their data, 'normalised' in a uniform format. In other words, a personal data cloud could be compared to one's Dropbox folder or a digital wallet where a master copy of personal data is stored. Every personal data cloud is supported by software that helps the user keep the data organised and understandable to an ordinary person. Crucially, the personal data cloud can only be accessed by the individual consumer herself; third parties can access specific data fragments only with the individual's prior authorisation. Then individuals can download applications created by third-party brands and developers that help extract everyday value from the various data sources.[75]

Wearable devices such as the Fitbit activity tracker, Oura ring, or an Apple watch, provide a paradigmatic use case of this model. Wearables are one of the most rapidly developing areas where the opportunities and risks, and interaction with personal data become most apparent.[76] Many early versions of wearable devices such as smartwatches or

---

[73] Paulius Jurcys, 'Liberty. Equality. Data. Podcast with Ann Cavoukian: Levelling the Playing Field between Humans and Algorithms' (25 April 2021) available at:
<https://medium.com/prifina/liberty-equality-data-podcast-episode-5-9760d273e35> accessed 15 January 2022.
[74] Jurcys et al. (n7); see also, e.g., Corrado Moiso and Roberto Minerva, 'Towards a user-centric personal data ecosystem: The role of the bank of individuals' data' (2012) 16th International Conference on Intelligence in Next Generation Networks 202; Donald McCarthy, Paul Malone, Johannes Hange, Kenny Doyle, Eric Robson, Dylan Conway et al., 'Personal Cloudlets: Implementing a User-centric Datastore with Privacy Aware Access Control for Cloud-Based Data Platforms' (2015) 2015 IEEE/ACM 1st International Workshop on Technical and Legal Aspects of Data Privacy and Security 38.
[75] Jurcys et al. (n7).
[76] Mahmoud Barhamgi, Charith Perera, Chirine Ghedira, and Djamal Benslimane, 'User-centric Privacy Engineering for the Internet of Things' (2018) IEEE Cloud Computing (September/October) 47; Marcus A Banks, 'Tech giants, armed with wearables data, are entrenching in health research' (2020) 24(4-5) Nature Medicine; Patrick Seitz, 'How Digital Health Firms Are Innovating Far Beyond Your Wrist' *Investor's Business Daily* (6 January 2022) available at:
m<https://www.investors.com/news/technology/digital-health-firms-advance-wearable-tech-at-ces-2022/> accessed 15 January 2022; 'How wearable health trackers could disrupt medicine' *The Economist* (11 January smart rings have been criticised for relying on inaccurate data.[77] Nevertheless, many old and new players in the wearables market are making significant progress in the accuracy of data collected from sensors integrated into intelligent wearable devices. For example, in October 2021, Oura released the third generation of its smart ring, which they claim to be the market's most accurate and precise smart ring.[78] Other hardware technology makers are coming with new sensors whose accuracy meets the highest thresholds previously only associated with medical devices.[79]

Given the increasing amount of data generated by wearables, how can the value from such data be captured on the users' side? The user-held data model provides a compelling approach to several practical questions about wearables data. At what point in time would such data become relevant, usable, and valuable? Who owns such data: the device manufacturer or the individual who bought the device?[80] What are the optimal ways to generate value from such data?[81] What if the device manufacturer is not creative enough to offer effective feedback to the device user? Considering that the amount of data generated is exponentially increasing,[82] such questions spark much debate from technological, legal, and ethical perspectives.[83] As such, the user-held data model solves many complex GDPR compliance-related issues that businesses, developers, and individuals face. The user-held

---

2022) available at:
<https://www.economist.com/podcasts/2022/01/11/how-wearable-health-trackers-could-disrupt-medicine>
accessed 15 January 2022.

[77] Lisa Eadicicco, 'Fitbit and Apple know their smartwatches aren't medical devices. But do you?' *Cnet*
(14 January 2022) available at:
<https://www.cnet.com/google-amp/news/fitbit-apple-know-smartwatches-arent-medical-devices-but-do-you/>
accessed 15 January 2022.

[78] cf Oura, available at: <ouraring.com> accessed 15 January 2022.

[79] Jessica Rendall, 'Wearable sensors that track glucose, ketones and alcohol levels are the future' *Cnet*
(13 January 2022) available at:
<https://www.cnet.com/google-amp/news/wearable-sensors-that-track-glucose-ketones-and-alcohol-levels-are-the-future/> accessed 15 January 2022.

[80] Josef Drexl, 'Connected Devices – an Unfair Competition Law Approach to Data Access Rights of Users' in German Federal Ministry of Justice and Consumer Protection, Max Planck Institute for Innovation and Competition (eds) *Data Access, Consumer Interests and Public Welfare* (Nomos, 2021) 477-527; Paulius Jurcys, 'The Proposed EU Data Act: 10 Key Takeaways' *Medium* (23 March 2022) available at:
<https://medium.com/prifina/the-proposed-eu-data-act-10-key-takeaways-6a380303c4f0> accessed 7 April 2022.

[81] cf n61.

[82] cf EU Commission's estimates, available at:
<https://ec.europa.eu/info/strategy/priorities-2019-2024/europe-fit-digital-age/european-data-strategy_en>
accessed 15 January 2022.

[83] Mark Fenwick and Paulius Jurcys, 'From Cyborgs to Quantified Selves: Augmenting Privacy Rights with User-Centric Technology and Design' (2022) 13(1) JIPITEC 72.

data model also unlocks user-generated data from siloes and offers a new approach to data which helps avoid risks related to data transfers.

In a user-held data model, the individual must first connect various data sources to the personal data cloud. This is done via a simple dashboard that allows users to quickly pick and choose the devices and data sources they own and connect to their personal data clouds. In the background, such connection of multiple data sources is made possible through complex APIs. Many wearable manufacturers such as Fitbit or Oura already provide APIs[84] to help connect such data sources to the individual's personal data cloud. Second, special software embedded in the personal data cloud takes the data from different devices and normalises it to make it coherent and uniform. The user can then see a custom-tailored overview of such data in the personal data dashboard.[85]

Individuals have full ownership and control over such combined data in their personal data cloud.[86] They also have full autonomy to decide how their data is used in the personal data cloud.[87] However, what is the actual value for individuals from having their data in their personal data cloud? While the possibility of connecting and pulling data from different data sources could be conceptually appealing, nowadays, consumers have very few tools to get value from the data they generate.

The fundamental proposition of the user-held data model is to create an ecosystem where the value from personal data is captured on the users' side. This can be achieved if individuals can activate their data in ways that provide meaningful everyday insights, recommendations, and other sorts of valuable utilities from such user-held data. Such value can be created by third-party developers and brands who build new applications that help unlock value from such data.[88] Suppose an individual has multiple wearable devices: an iWatch, Oura Ring, and other intelligent sensors at home, such as the Fitbit scales. What insights could be created if there was a way to combine data from all those devices? What about other data sources, such as the user's location data from Google Maps, credit card payments history, and other digital footprints on the user's favourite platforms, such as Netflix or Twitch? And what about public data (weather forecast, highway traffic, IMDB)? Myriads of applications could be created by correlating the data from various user-held and public data sources.[89]

---

[84] Jouko Ahvenainen, 'API-First: Winners Take It From Good Intentions to Real Action' *Medium* (29 October 2021) available at:
<https://medium.com/prifina/api-first-winners-take-it-from-good-intentions-to-real-action-7caa27efcc10> accessed 15 January 2022.
[85] Jurcys et al. (n7); Fenwick/Jurcys (n83) 73-77.
[86] cf Jurcys et al. (n 7).
[87] Paulius Jurcys, Chris Donewald, Jure Globocnik and Markus Lampinen, 'My Data, My Terms: A Proposal for Personal Data Use Licenses' (2020) 33 Harv JL & Tech Digest 9-12.
[88] Unlocking of data from silos and creation of new types of services on top of user-generated data seems to be among the main objectives of the EU Data Strategy and the Proposed EU Data Act, see Art. 35 of the EU Data Act and Jurcys (n80).

Access to data is one of the biggest challenges for developers and device makers. For example, a maker of a glucose monitoring sensor can collect biomarker data from that sensor and build an application that could offer some insights and recommendations to the user. However, the sensor maker's possibilities are limited to the data accessible from that specific device. Intelligent device manufacturers need access to other data sources to create new correlations of various data sets and build applications based on such data. In short, a user-held data model is an ecosystem-wide solution that aims to generate value on the user's side by:

- Creating an easy way for individuals to collect their data sources in their personal data clouds and assert full ownership and control over such data.
- Creating a unified standard for such user-held data.
- Creating open-source tools for developers to build applications that run on top of user-held data.

Capturing value on the user's side means that individuals can run such third-party apps *within* their personal data cloud: the data is processed locally (within the cloud), and the value stays on the user's side. Third-party brands and developers benefit from such an ecosystem because they can quickly build applications that in principle do not require back-end data solutions (the data is already unified within the user's personal data cloud), and third-party apps could be published and integrated into the user-held data ecosystem through public APIs.

---

[89] Jouko Ahvenainen, 'The Next Best Tools And Areas For Gold Mining' *Medium* (15 October 2021) available at: <https://medium.com/prifina/the-next-best-tools-and-areas-for-gold-mining-572ae2c0a0c6> accessed 15 January 2022.

**3.2 Cross-Border Data Transfers in a User-Held Data Environment**

What, then, do cross-border data transfers look like in a user-held data environment? From a legal point of view, a user-held data model is embedded in the fundamental principles of the GDPR and similar legislation elsewhere, such as the CCPA or the proposed EU Data Act. The individual should have the ultimate authority to control how 'their' data is used and with whom that data is shared.[90]

As such, by default, third-party application developers or brands do not have access to any data processed by their apps in the users' personal data clouds. In some instances, individuals may give access to anonymised profile cards that reflect, e.g., their age group, income level, or specific activities and habits. However, in such instances, third-party app developers and brands are viewing such data only with an individual's prior explicit consent and under the terms determined by the individual. The individual can exercise such control by granting permissions to access data on a granular level: data access permissions can have many different layers (e.g., access for a specific period, a particular purpose/function, or a specific third party).[91]

From an application developer's perspective, offering services that run 'on top' of the user-held data not only secures that the data is used lawfully, but it also means that the individual can get bespoke services under the terms which the individual herself sets (Art. 5(1)(a) GDPR). Accessing data directly from the individual aligns with the GDPR's principles of data minimisation (Art. 5(1)(c) GDPR) because the app developer does not have to hold any data on its servers. Rather, the application processes data in an individual's personal data cloud or on a device.

The principle of purpose limitation is also upheld because the data is used only for the specific purpose to which the individual granted prior permission (Art. 5(1)(b) GDPR).

A similar claim can be made about the principle of data portability. As indicated above, in the user-held data environment, third-party applications are installed and run 'on top of' the user-held data. Hence, by default, these apps can access only that specific segment of an individual's personal data, necessary to perform the specified functions of the application. Furthermore, such data is processed locally (either in the user's personal data cloud or an edge device). From that perspective, the user-held data model clarifies the practical implementation of the data portability principle envisioned in the GDPR. Rather than bringing data from service provider 'A' to service provider 'B', service providers are prompted to come directly to the user. This avoids creating unnecessary copies of personal data with third parties and ascertains that the value from the user's personal data is captured on the user's side.

---

[90] cf Paul De Hert et al. (n60).
[91] cf Jurcys et al. (n87).

As such, the problems related to cross-border data transfers are greatly reduced in the user-held data model because, in principle, all the data is processed locally (within the user's personal data cloud or on the user's edge device). There is no data exporter or data importer. The factors considered in making data transfer impact assessments (e.g., assessing the risk of prohibited lawful access in the target jurisdiction or the need to identify safeguards in place) are practically irrelevant.

The only relevant cross-border consideration could be related to the physical location of servers in which the individual's personal data cloud and the data are held. However, in the user-held data environment, the individual can – if they wish to – select: (i) the service provider where the data and the personal data cloud is stored (e.g., Amazon, Google Drive, or some local service provider in the state where the individual is habitually resident), and (ii) individuals could further choose a geographical location where their personal data cloud and user-generated data are held (or move that data from one state to another).

Let us illustrate the above by looking at two practical examples related to sharing the health and wellness data collected by an individual (Tom) from three personal wearable devices: Apple Watch, Oura Ring, and Fitbit Scales. Suppose that Tom connected those three devices to his personal data cloud, and the data is now constantly collected from those three data sources. Tom's personal data cloud contains a 'software robot,' which makes that data unified and standardised.

- *Use-case No. 1: Local Data Apps.* Tom can download various local data apps that help him get better insights from the data collected by his three devices. For example, there may be an app (we can call it 'BetterSleep') that helps correlate Tom's sleep data from his Oura ring (latency, resting heart rate, sleep hours) with daily activity data from the Apple watch (number of steps taken, and minutes of exercise, heart

Electronic copy available at: https://ssrn.com/abstract=4010356

rate). The BetterSleep app helps Tom understand how his daily activity affects his sleep, offers insights into his daily routines, and recommends ways to improve his habits to enhance physical and mental health. From the GDPR perspective, the BetterSleep app is installed locally (in Tom's personal data cloud) and processes Tom's data there; the publisher of the BetterSleep app does not have access to the data that the app processes Tom's data within Tom's personal data, thus negating the need for any data transfer in this use case.

- *Use Case No. 2: Sharing Health and Wellness Data.* Suppose that Tom works at a hospital as a nurse. As the number of hospitalised persons increases, Tom has to work long 'after-hours.' Suppose that the hospital managers are legally required to check the health status of their employees regularly (e.g., to detect COVID symptoms and fatigue level). Suppose that the hospital managers elect to create an 'HealthyMind' application which its staff members could choose to download on their phones. The MealthyMind app is intended to collect anonymized reports about hospital staff members' health and wellness status based on the data collected from the employees' wearable devices. The HealthyMind app may have pre-set parameters of data that it needs to create a report about Tom's current physical state (e.g., based on body temperature, heart rate, rest status, etc.). Tom could set such a report to be sent to the hospital at designated hours of the day; but the hospital managers only see a 'green' icon if Tom's health data meets the threshold; 'yellow' or 'red' colours in cases where some parameters are outside of the normal range. Such reports could be sent in an end-to-end encrypted fashion. The report here is a concept of data sharing. Consider alternatives: (a) the processing happens in Tom's environment, and any result is only shown to Tom. The hospital sees nothing (i.e., it is truly private), or (b) the same as above, but only an outlier value is shared, with the hospital but then it is never linked to Tom, but rather shared with the hospital as '1 staff member had an outlier value.' The hospital could also offer suggestions and have those displayed to Tom, without them identifying Tom specifically to the hospital. Again, such reports are generated within the user's (in this case, Tom's) personal data cloud, and thus negate the need for data transfers.

In this way, legal risk is mitigated – by greatly reducing the need for cross-border transfers of data and the user experience is improved – by giving users greater control over 'their' personal data.

## 4 Towards a New Data Economy

Transnational data transfers are one of the most complex issues in the realm of data privacy. But as this paper has suggested, it is also an area where major changes are likely to occur in the medium-term because of the new opportunities that will be unlocked by emerging technologies.[92] Just as the ad businesses are preparing for a new reality without third-party cookies,[93] technology companies should be open to exploring new data collection and processing approaches that are emerging in the market.

The user-held data model introduced above is not the only solution to quandaries that permeate cross-border data transfers and the new data economy. In the EU, for example, governments and the private sector are working on Gaia-X – a new proposal for the next generation of data infrastructure: 'an open, transparent and secure digital ecosystem, where data and services can be made available, collated and shared in an environment of trust.'[94] Rapidly emerging technologies are challenging the currently prevailing data collection and processing practices and we will see further technological breakthroughs in the next few years. Crucially, these technologies are driven, in part, by regulatory developments and designed with the goal of compliance and mitigating legal risk, as well as delivering a better user experience that allows individuals, as well as businesses, to extract value from their personal data.

---

[92] Yuko Suda, *The Politics of Data Transfer: Transatlantic Conflict And Cooperation Over Data Privacy*, Routledge 2018) 1.
[93] Tim Glomb, 'Say Goodbye to Cookies' *HBR* (8 April 2021) available at:
<https://hbr.org/2021/04/say-goodbye-to-cookies> accessed 15 January 2022.
[94] 'What is Gaia-X?' available at:
<https://www.data-infrastructure.eu/GAIAX/Navigation/EN/Home/home.html> accessed 15 January 2022.